\documentclass[conference]{IEEEtran}
\IEEEoverridecommandlockouts
\usepackage{cite}
\usepackage{amsmath,amssymb,amsfonts}
\usepackage{algorithmic}
\usepackage{graphicx}
\usepackage{textcomp}
\usepackage{xcolor}
\usepackage{balance}

\newtheorem{remark}{Remark}
\newtheorem{theorem}{Theorem}

\newtheorem{corollary}{Corollary}
\newtheorem{definition}{Definition}

\def\BibTeX{{\rm B\kern-.05em{\sc i\kern-.025em b}\kern-.08em
    T\kern-.1667em\lower.7ex\hbox{E}\kern-.125emX}}

\begin{document}

\title{Energy Efficiency Analysis of Intelligent Reflecting Surface System with Hardware Impairments\\
}

\author{\IEEEauthorblockN{Yiming Liu, Erwu Liu and Rui Wang}
\IEEEauthorblockA{\textit{College of Electronics and Information Engineering, Tongji University, Shanghai, China} \\
Emails: 15995086362@163.com, erwu.liu@ieee.org, ruiwang@tongji.edu.cn}
}

\maketitle

\begin{abstract}
Recently, as explosive growth of mobile data traffic, the performance of wireless communication systems requires to be enhanced significantly in future. Intelligent reflecting surface (IRS) can be used as a promising way to improve the energy efficiency of wireless communications with less complexity and hardware cost. Most existing studies consider the cases with ideal hardware, however, both physical transceiver and IRS suffer from non-negligible hardware impairments which may greatly degrade the system performance. In this paper, by taking hardware impairments into consideration, we focus on the energy efficiency analysis of IRS system. Our first contribution is to derive the optimal receive combining and transmit beamforming vectors. After that, we characterize the asymptotic channel capacity. With the derived asymptotic channel capacity and the power consumption model, the analytical upper and lower bounds of energy efficiency are provided. Our results show that an IRS system can achieve both high spectral efficiency and high energy efficiency with moderate number of antennas. This observation is encouraging for that there is no need to cost a lot on expensive high-quality antennas, which corresponds to the requirements of new communication paradigms.
\end{abstract}


\section{Introduction}
The March 2020 report, which was released by Cisco Systems, Inc., shows that the number of networked devices and connections will reach up to 29.3 billions by the year of 2023, and about half of them are mobile-ready devices and connections \cite{index2019global}. With the growth of mobile-ready devices and connections, there exists an explosive growth of mobile data traffic. The 5th generation (5G) wireless network technology has been standardized to solve these problems. However, there is no single enabling technology that can support all 5G application requirements during the standardization process \cite{8796365}. Some new use cases will bring more challenging communication engineering problems, which necessitates radically new communication paradigms, especially at the physical layer. Intelligent reflecting surface (IRS) has emerged as a new solution to improve the energy and spectrum efficiency of wireless communication systems. Prior works have revealed that IRS can effectively control the wavefront,\textit{ e.g.}, the phase, amplitude, frequency, and even polarization, of the impinging signals without the need of complex decoding, encoding, and radio frequency processing operations \cite{8796365,8930608,8936989,8647620}.

Existing studies usually consider the cases with ideal hardware of transceiver and IRS, and their results rely on asymptotics. However, both physical transceiver and IRS suffer from non-negligible hardware impairments which may greatly degrade the performance of systems. It has been shown that hardware impairments bound the channel capacity and energy efficiency in massive multiple-input multiple-output (MIMO) system as the signal-to-noise ratio (SNR) approaches infinity \cite{6362131}. Energy efficiency is an important metric in evaluating the performance of communication systems. It is well known that multiple-antenna technique can offer improved power efficiency, owing to both array gains and diversity effects. The power transmitted by the user can be cut inversely proportional to the square-root of the number of base station (BS) antennas with no reduction in performance \cite{6457363}. However, the requirements of high hardware cost and high complexity are still the main hindrances to its implementation. The recent advent of IRS-assisted wireless communication system can be used as an alternative solution to improve the energy and spectrum efficiency with less complexity and hardware cost. 

Thus, in this paper, by taking the hardware impairments into account, we analyze the energy efficiency of the IRS-assisted wireless communication system. Our first contribution is to derive the optimal receive combining and transmit beamforming vectors. After that, the asymptotic channel capacity is characterized. We consider two types of asymptotics, that is, the infinite transmit power, and the infinite  numbers of antennas and reflecting elements. Our results show that the hardware impairments bound the channel capacity, and the major function of IRS is that the channel capacity limit can be achieved at low transmit power, while the IRS cannot increase the asymptotic channel capacity. Finally, with the derived asymptotic channel capacity limits and the power consumption model, we analyze the energy efficiency and present the upper and lower bounds. Our results show that hardware impairments set a limit for energy efficiency, and IRS offers a significant improvement in energy efficiency when the number of BS antennas is fixed. In addition, the obtained asymptotic channel capacity and energy efficiency show that an IRS system can achieve both high spectral efficiency and high energy efficiency without the need of costing a lot on BS antennas.

\section{Communication System Model}
\begin{figure}[htbp]
	\centerline{\includegraphics[width=8.7cm]{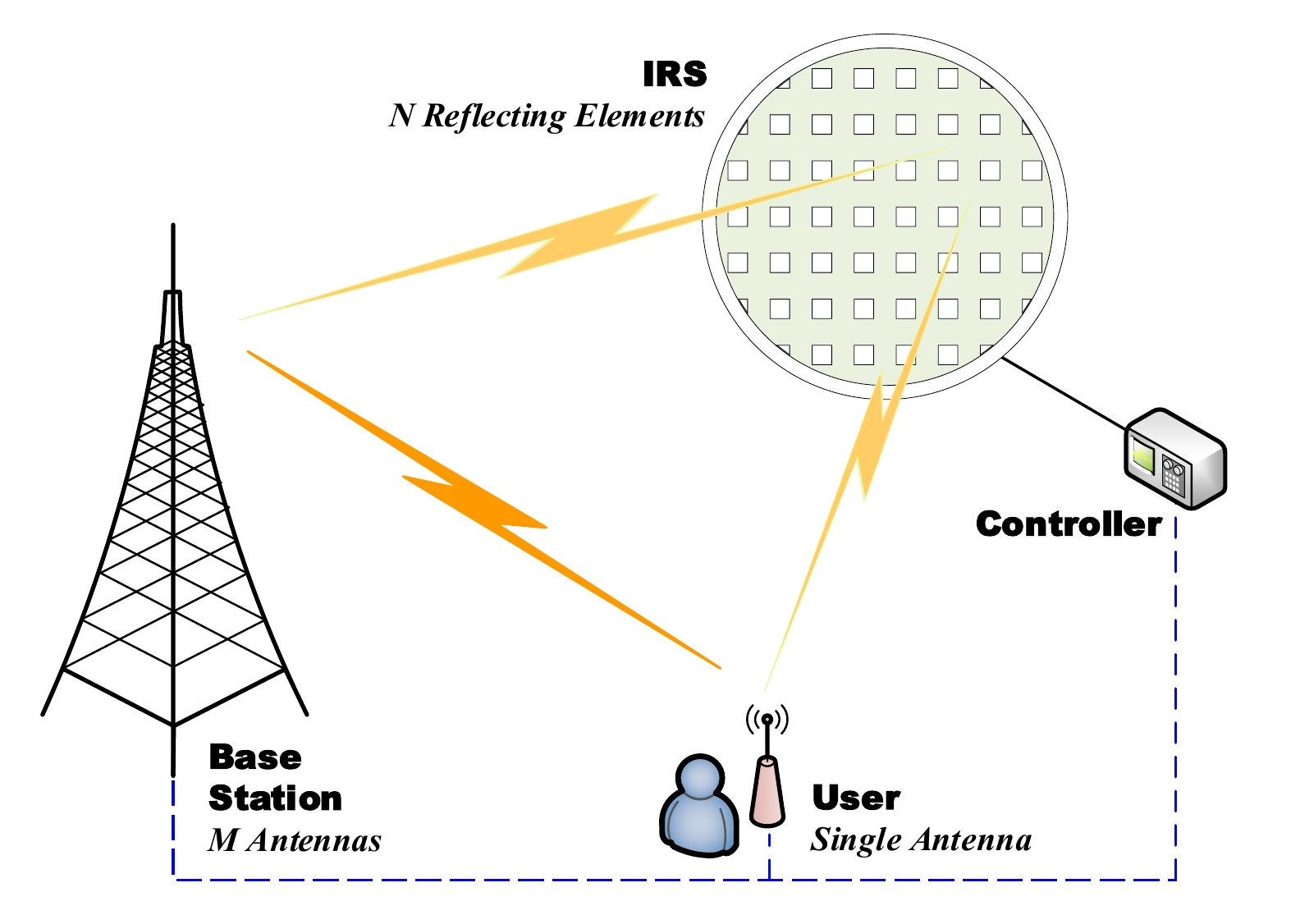}}
	\caption{The IRS-assisted wireless communication system with an $M$-antenna BS, a single-antenna user, and an IRS comprising $N$ reflecting elements.}
	\label{fig1}
	\vspace{-0.4 em}
\end{figure}

We consider an IRS-assisted wireless communication system, as illustrated in Fig.~\ref{fig1}. The system consists of an $ M $-antenna BS, a single-antenna user, and an IRS comprising $ N $ reflecting elements. Based on the physically correct system models in prior works \cite{8936989, 8930608, 8647620}, we give the communication system model in this section. The operations at the IRS is represented by the diagonal matrix $ \mathbf{\Phi} = \operatorname{diag} \left( e^{j\theta_{1}},\cdots, e^{j\theta_{N}} \right) $, where $ \theta_{i} \in [0,2\pi] $ represents the phase-shift of the $ i^{th} $ reflecting element. The channel realizations are generated randomly and are independent between blocks, which basically covers all physical channel distributions. Denote the channels of BS-user link, BS-IRS link and IRS-user link as $ \mathbf{h}_{\mathrm{d}} \in \mathbb{C}^{M \times 1} $, $ \mathbf{G} \in \mathbb{C}^{M \times N} $ and $ \mathbf{h}_{\mathrm{r}} \in \mathbb{C}^{N \times 1} $, respectively. They are modeled as ergodic processes with fixed independent realizations, $ \mathbf{h}_{\mathrm{d}} \sim \mathcal{C}\mathcal{N} \left( 0, \mathbf{C}_{\mathrm{d}} \right) $ and $ \mathbf{H}_{\mathrm{IRS}}=\mathbf{G}\operatorname{diag}\left(\mathbf{h}_{\mathrm{r}}\right) \sim \mathcal{C}\mathcal{N}\left(0,\mathbf{C}_{\mathrm{IRS}}\right) $, where $\mathcal{C}\mathcal{N}(\cdot)$ represents a circularly symmetric complex Gaussian distribution, and $\mathbf{C}_{\mathrm{d}}$, $\mathbf{C}_{\mathrm{IRS}}$ are the positive semi-definite covariance matrices.

\vspace{-1.2 em}

\begin{figure}[htbp]
	\centerline{\includegraphics[width=7.5 cm]{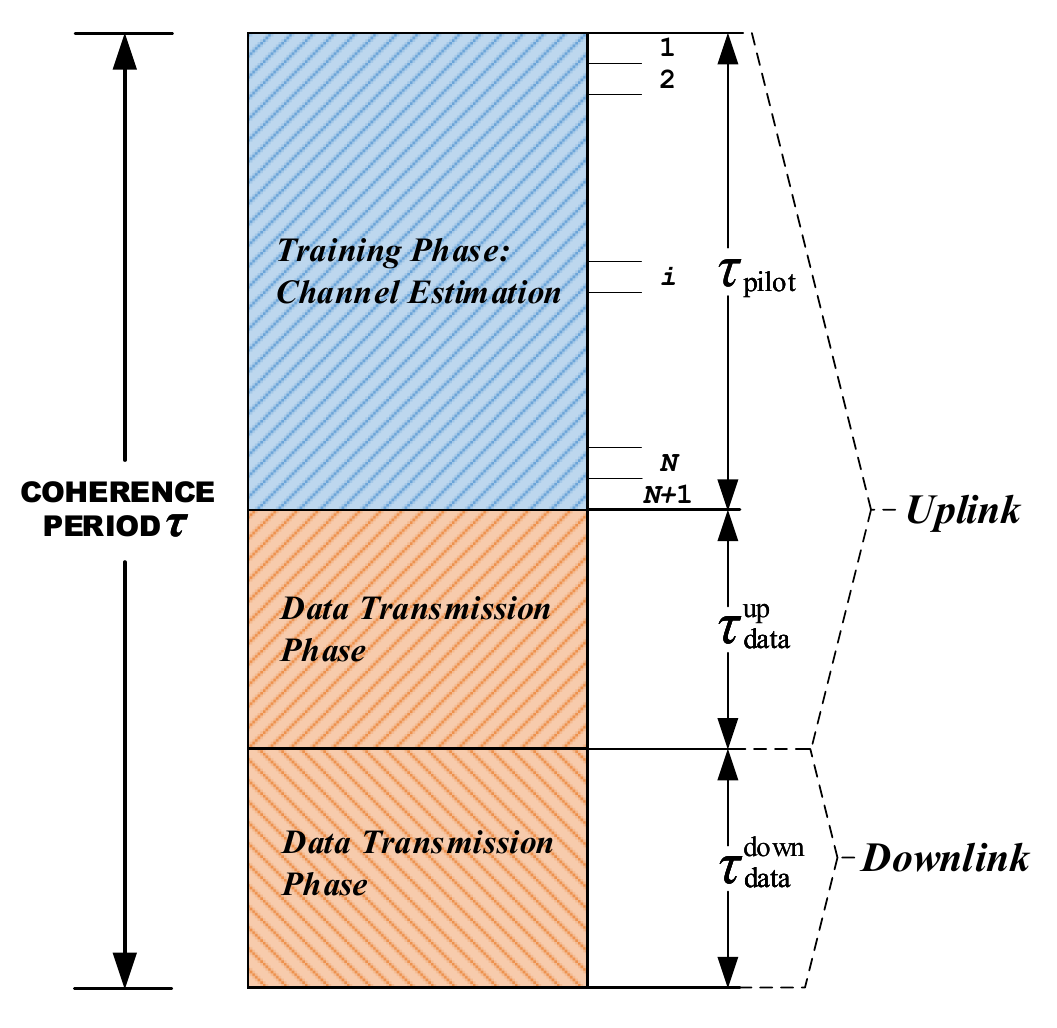}}
	\caption{The communication protocol that we adopt for the IRS-assisted wireless communication system.}
	\label{fig2}
	\vspace{-1.25 em}
\end{figure}

We adopt the protocol proposed in \cite{nadeem2019intelligent} for the IRS system, as illustrated in Fig.~\ref{fig2}. The channel coherence period $ \tau $ is divided into an uplink training phase of $ \tau_{\mathrm{pilot}} $, an uplink transmission phase of $ \tau_{\mathrm{data}}^{\mathrm{up}} $ and a downlink transmission phase of $ \tau_{\mathrm{data}}^{\mathrm{down}} $. During the uplink training phase, the user transmits deterministic pilot signal $ x $ to estimate channels, where the average power of $ x $ is $ \mathbb{E}\left\{|x|^{2}\right\} = p_{\mathrm{UE}} $. Since the IRS is a passive device, the BS has to estimate both channels $\mathbf{G}$ and $\mathbf{h}_{\mathrm{r}}$, where $\mathbf{G}$ and $\mathbf{h}_{\mathrm{r}}$ are cascaded as $ \mathbf{H}_{\mathrm{IRS}} = \mathbf{G}\operatorname{diag}\left(\mathbf{h}_{\mathrm{r}}\right) = \left[\mathbf{h}_{1}, \cdots, \mathbf{h}_{N}\right] $. Each column vector $\mathbf{h}_{i} \sim \mathcal{C}\mathcal{N}\left(0,\mathbf{C}_{i}\right)$ in $ \mathbf{H}_{\mathrm{IRS}} $ represents the channel between the BS and the user through IRS when only the $ i^{th} $ reflecting element is ON. The uplink training phase is divided into $ N+1 $ subphases. During the $ 1^{st} $ subphase, all reflecting elements are OFF and the BS estimates the direct channel $ \mathbf{h}_{\mathrm{d}} $; During the $ (i+1)^{th} $ subphase, only the $ i^{th} $ reflecting element is ON and the BS estimates the channel $ \mathbf{h}_{i} $. By exploiting channel reciprocity, the BS will transmit data to the user during the downlink transmission phase.

The aggregate hardware impairments of transceiver can be modeled as independent additive distortion noises \cite{5456453, zetterberg2011experimental}.
The distortion noise at the user $\eta_{\mathrm{UE}} \in \mathbb{C}$ obeys the distribution of $\mathcal{C} \mathcal{N}\left(0, v_{\mathrm{UE}}\right)$, and the distortion noise at the BS $\boldsymbol{\eta}_{\mathrm{BS}} \in \mathbb{C}^{M\times1} $ obeys the distribution of $\mathcal{C N}\left(\mathbf{0}, \mathbf{\Upsilon}_{\mathrm{BS}}\right)$, where $v_{\mathrm{UE}}$ and $\mathbf{\Upsilon}_{\mathrm{BS}}$ are the variance/covariance matrix of the distortion noise. The distortion noise at an antenna is proportional to the signal power at this antenna \cite{5456453, zetterberg2011experimental}, thus we have:
\begin{itemize}
	\item During the downlink data transmission phase, $\mathbf{\Upsilon}_{\mathrm{BS}}$ can be modeled as $\mathbf{\Upsilon}_{\mathrm{BS}}=\kappa_{\mathrm{BS}} \operatorname{diag}\left(\mathbf{Q}\right)$, where the matrix $\mathbf{Q}$ represents the covariance matrix of the transmitted data signal $\mathbf{x} \in \mathbb{C}^{M\times1}$, $\mathbf{Q}=\mathbb{E}\left\{\mathbf{x}\mathbf{x}^{\mathrm{H}}\right\}$. $v_{\mathrm{UE}}$ can be modeled as $v_{\mathrm{UE}}=\kappa_{\mathrm{UE}}\mathbf{h}^{\mathrm{H}}\left(\mathbf{x}+\boldsymbol{\eta}_{\mathrm{BS}}\right)\left(\mathbf{x}+\boldsymbol{\eta}_{\mathrm{BS}}\right)^{\mathrm{H}}\mathbf{h}$, where $\mathbf{h}$ represents the overall channel $\mathbf{h}_{\mathrm{d}}+\mathbf{G}\mathbf{\Phi}\mathbf{h}_{\mathrm{r}}$ for simplicity.	$\kappa_{\mathrm{BS}}$ and $\kappa_{\mathrm{UE}}$ are respectively the proportionality coefficients which characterize the levels of impairments at the BS and the user, and are related to the error vector magnitude (EVM). The EVM is a common measure of hardware quality for transceivers, \textit{e.g.}, when the BS transmits the signal $\mathbf{x}$, the EVM at the BS is
	\begin{equation}
	\mathrm{EVM}_{\mathrm{BS}} = \sqrt{\frac{\operatorname{tr}\left( \mathbb{E}\left\{\boldsymbol{\eta}_{\mathrm{BS}} \boldsymbol{\eta}_{\mathrm{BS}}^{\mathrm{H}}\right\} \right) }{\operatorname{tr}\left( \mathbb{E}\left\{\mathbf{x} \mathbf{x}^{\mathrm{H}}\right\}\right) }} = \sqrt{\kappa_{\mathrm{BS}}}.
	\end{equation}
	\item During the uplink data transmission phase, $ v_{\mathrm{UE}} $ can be modeled as $v_{\mathrm{UE}}=\kappa_{\mathrm{UE}} p_{\mathrm{UE}}$, and $ \mathbf{\Upsilon}_{\mathrm{BS}} $ can be modeled as $\mathbf{\Upsilon}_{\mathrm{BS}}=\kappa_{\mathrm{BS}}\left(p_{\mathrm{UE}}+\kappa_{\mathrm{UE}}p_{\mathrm{UE}}\right)\operatorname{diag}\left(\mathbf{C}_{\mathrm{d}}+\sum_{i=1}^{N}\mathbf{C}_{i}\right)$.
\end{itemize}

The hardware impairments of IRS can be modeled as phase noise since the IRS is a passive device and high-precision configuration of the reflection phases is infeasible. The phase noise of the $i^{th}$ element of IRS is denoted as $\Delta \theta_{i}$, which is randomly distributed on $[-\pi, \pi)$ according to a certain circular distribution. Due to the reasonable assumption in \cite{8869792}, the distribution of the phase noise $\Delta \theta_{i}$ has mean direction zero ($\arg \left(\mathbb{E}\left\{e^{j \Delta \theta_{i}}\right\}\right)=0$) and its probability density function is symmetric around zero. The actual matrix of IRS with phase noise is $\widetilde{\mathbf{\Phi}}=\operatorname{diag}\left(e^{j(\theta_{1}+\Delta\theta_{1})}, e^{j(\theta_{2}+\Delta\theta_{2})}, \cdots, e^{j(\theta_{N}+\Delta\theta_{N}}\right)$.

Based on the communication system model given above, the received signal $\mathbf{y}\in\mathbb{C}^{M\times1}$ at the BS during the uplink transmission phase from the user is
\begin{equation}
\label{eq.2}
\mathbf{y}=\left(\mathbf{h}_{\mathrm{d}}+\mathbf{G}\widetilde{\mathbf{\Phi}}\mathbf{h}_{\mathrm{r}}\right)\left(x+\eta_{\mathrm{UE}}\right)+\boldsymbol{\eta}_{\mathrm{BS}}+\mathbf{n},
\end{equation}
where $x \in \mathbb{C}$ is the transmitted data signal, and $\mathbf{n} \in \mathbb{C}^{M \times 1}$ is an additive white Gaussian noise with the elements independently drawn from $\mathcal{C} \mathcal{N}\left(0, \sigma_{\mathrm{BS}}^{2}\right)$. The transmit power of the user is $ p_{\mathrm{UE}} = \mathbb{E}\left\{|x|^{2}\right\} $.
The received signal $y\in\mathbb{C}$ at the user during the downlink transmission phase from the BS is
\begin{equation}
\label{eq.3}
y=\left(\mathbf{h}_{\mathrm{d}}^{\mathrm{H}}+\mathbf{h}_{\mathrm{r}}^{\mathrm{H}} {\widetilde{\mathbf{\Phi}}}^{\mathrm{H}} \mathbf{G}^{\mathrm{H}}\right)\left(\mathbf{x}+\boldsymbol{\eta}_{\mathrm{BS}}\right)+\eta_{\mathrm{UE}}+n,
\end{equation}
where $ \mathbf{x}\in\mathbb{C}^{M \times 1} $ is the transmitted data signal and $n\in\mathbb{C}$ is an additive white Gaussian noise drawn from $\mathcal{C}\mathcal{N}\left(0, \sigma_{\mathrm{UE}}^{2}\right)$. The transmit power of the BS is $ p_{\mathrm{BS}} = \mathbb{E}\left\{\mathbf{x}^{\mathrm{H}}\mathbf{x}\right\} $.

\section{Channel Capacity and Energy Efficiency}

\subsection{Asymptotic Channel Capacity}
In this subsection, we analyze the channel capacity of the uplink and downlink in Eqs. (\ref{eq.2}) and (\ref{eq.3}). In each coherence period $ \tau $, the BS has the imperfect channel state information $ \mathcal{H}_{\mathrm{BS}} $ of the actual channel states $ \mathcal{H} $ by using the channel estimation method, \textit{e.g.}, linear minimum mean square error estimator, and the user has the imperfect channel state information $ \mathcal{H}_{\mathrm{UE}} $. The uplink and downlink channel capacity can be expressed as
\begin{equation}
{\mathrm{C}}_{\mathrm{up}}=\frac{\tau_{\mathrm{data}}^{\mathrm{up}}}{\tau} \mathbb{E}\left\{\max \; \mathcal{I}\left(x ; \mathbf{y} | \mathcal{H}, \mathcal{H}_{\mathrm{BS}}, \mathcal{H}_{\mathrm{UE}}\right)\right\},
\end{equation}		
\begin{equation}
\label{eq.5}
{\mathrm{C}}_{\mathrm{down}}=\frac{\tau_{\mathrm{data}}^{\mathrm{down}}}{\tau} \mathbb{E}\left\{\max \; \mathcal{I}\left(\mathbf{x} ; y | \mathcal{H}, \mathcal{H}_{\mathrm{BS}}, \mathcal{H}_{\mathrm{UE}}\right)\right\},
\end{equation}
where $ \mathcal{I}\left(x ; \mathbf{y} | \mathcal{H}, \mathcal{H}_{\mathrm{BS}}, \mathcal{H}_{\mathrm{UE}}\right) $ and $ \mathcal{I}\left(\mathbf{x} ; y | \mathcal{H}, \mathcal{H}_{\mathrm{BS}}, \mathcal{H}_{\mathrm{UE}}\right) $ are the conditional mutual information. 

For any given phase shifts of IRS, it can be verified that the maximum-ratio transmission (MRT) is the optimal transmit beamforming solution \cite{8683145,8930608}. Referring to \cite{6891254}, this conclusion can be easily extended to the case with hardware impairments, which is shown as follows.

\begin{theorem}
	To maximize the average SNR in the IRS-assisted wireless communication system with hardware impairments, the receive combining and transmit beamforming vectors are given by
	\begin{equation}
	\mathbf{w}_{\mathrm{r}}=\frac{\mathbf{U}^{-1} \widetilde{\mathbf{h}}^{\mathrm{H}}}{\left\|\mathbf{U}^{-1} \widetilde{\mathbf{h}}^{\mathrm{H}}\right\|_{2}},
	\end{equation} 
	\begin{equation}
	\label{71}
	\mathbf{w}_{\mathrm{t}}=\frac{\mathbf{D}^{-1} \widetilde{\mathbf{h}}}{\left\|\mathbf{D}^{-1} \widetilde{\mathbf{h}}\right\|_{2}},
	\end{equation} 
	where $ \mathbf{U}=\left(1+\kappa_{\mathrm{UE}}\right) \kappa_{\mathrm{BS}} \operatorname{diag}(\widetilde{\mathbf{h}}\widetilde{\mathbf{h}}^{\mathrm{H}})+\kappa_{\mathrm{UE}} \widetilde{\mathbf{h}} \widetilde{\mathbf{h}}^{\mathrm{H}} + \frac{\sigma_{\mathrm{UE}}^{2}}{p_{\mathrm{UE}}} \mathbf{I}$, $ \mathbf{D}=\left(1+\kappa_{\mathrm{UE}}\right) \kappa_{\mathrm{BS}} \operatorname{diag}(\widetilde{\mathbf{h}}\widetilde{\mathbf{h}}^{\mathrm{H}})+\kappa_{\mathrm{UE}} \widetilde{\mathbf{h}} \widetilde{\mathbf{h}}^{\mathrm{H}} + \frac{\sigma_{\mathrm{UE}}^{2}}{p_{\mathrm{BS}}} \mathbf{I}$, and $ \widetilde{\mathbf{h}}$ represents the overall channel $\mathbf{h}_{\mathrm{d}}+\mathbf{G}\widetilde{\mathbf{\Phi}}\mathbf{h}_{\mathrm{r}}$ for simplicity.
\end{theorem}
\begin{IEEEproof}
	Suppose that the phase shift parameters of IRS are optimal, the upper bound of downlink channel capacity in Eq. (\ref{eq.5}) is
	\begin{equation}
	\label{81}
	\mathrm{C}_{\mathrm{down}} \leq \frac{\tau_{\mathrm{data}}^{\mathrm{down}}}{\tau} \mathbb{E}\left\{\max _{\|\mathbf{w}_{\mathrm{t}}\|_{2}=1} \log _{2}(1+\operatorname{SNR}(\mathbf{w}_{\mathrm{t}}))\right\},
	\end{equation}
	where
	\begin{equation}
	\label{eq.91}
	\operatorname{SNR}(\mathbf{w}_{\mathrm{t}})=\frac{|\widetilde{\mathbf{h}}^{\mathrm{H}}\mathbf{x}|^{2}} {\widetilde{\mathbf{h}}^{\mathrm{H}} \mathbf{\Upsilon}_{\mathrm{BS}} \widetilde{\mathbf{h}} + v_{\mathrm{UE}} + \sigma_{\mathrm{UE}}^{2} }.
	\end{equation}
	 The transmit beamforming vector $\mathbf{w}_{\mathrm{t}}$ is in unit-form, and the transmitted data signal $ \mathbf{x}$ has the form of $\mathbf{x} = \mathbf{w}_{\mathrm{t}}x$. According to the distortion noise modeled in Section II, $\widetilde{\mathbf{h}}^{\mathrm{H}} \mathbf{\Upsilon}_{\mathrm{BS}} \widetilde{\mathbf{h}}^{\mathrm{H}}$ in Eq. (\ref{eq.91}) can be rewritten as
	\begin{equation}
	\label{101}
	\begin{aligned}
	\widetilde{\mathbf{h}}^{\mathrm{H}} \mathbf{\Upsilon}_{\mathrm{BS}} \widetilde{\mathbf{h}}^{\mathrm{H}} &=\kappa_{\mathrm{BS}} \widetilde{\mathbf{h}}^{\mathrm{H}} \operatorname{diag}\left(\mathbf{Q}\right) \widetilde{\mathbf{h}} \\
	&=\kappa_{\mathrm{BS}} p_{\mathrm{BS}} \mathbf{w}_{\mathrm{t}}^{\mathrm{H}} \operatorname{diag}\left(\widetilde{\mathbf{h}}\widetilde{\mathbf{h}}^{\mathrm{H}}\right) \mathbf{w}_{\mathrm{t}}.
	\end{aligned}
	\end{equation}
	Similarly, $ v_{\mathrm{UE}} $ in Eq. (\ref{eq.91}) can be rewritten as 
	\begin{equation}
	\label{111}
	v_{\mathrm{UE}} = \kappa_{\mathrm{UE}} p_{\mathrm{BS}}\mathbf{w}_{\mathrm{t}}^{\mathrm{H}} \widetilde{\mathbf{h}} \widetilde{\mathbf{h}}^{\mathrm{H}} \mathbf{w}_{\mathrm{t}} + \kappa_{\mathrm{UE}} \widetilde{\mathbf{h}}^{\mathrm{H}} \mathbf{\Upsilon}_{\mathrm{BS}} \widetilde{\mathbf{h}}^{\mathrm{H}} .
	\end{equation}
	Then, by substituting Eqs. (\ref{101}) and (\ref{111}) into Eq. (\ref{eq.91}), we obtain
	\begin{equation}
	\label{121}
	\operatorname{SNR}(\mathbf{w}_{\mathrm{t}})=\frac{\mathbf{w}_{\mathrm{t}}^{\mathrm{H}} \widetilde{\mathbf{h}} \widetilde{\mathbf{h}}^{\mathrm{H}} \mathbf{w}_{\mathrm{t}}} {\mathbf{w}_{\mathrm{t}}^{\mathrm{H}} \mathbf{D} \mathbf{w}_{\mathrm{t}}},
	\end{equation}
	where $\mathbf{D}=\left(1+\kappa_{\mathrm{UE}}\right) \kappa_{\mathrm{BS}} \operatorname{diag}(\widetilde{\mathbf{h}}\widetilde{\mathbf{h}}^{\mathrm{H}})+\kappa_{\mathrm{UE}} \widetilde{\mathbf{h}} \widetilde{\mathbf{h}}^{\mathrm{H}} + \frac{\sigma_{\mathrm{UE}}^{2}}{p_{\mathrm{BS}}} \mathbf{I}$.
	The function $ \log _{2}(1+\operatorname{SNR}(\mathbf{w}_{\mathrm{t}}))$ in Eq. (\ref{81}) has the structure of $ f(x) = \log _{2}(1+x) $, and $ f(x) $ is a monotonically increasing function. Thus, the maximum of Eq. (\ref{81}) can be obtained by maximizing $\operatorname{SNR}(\mathbf{w}_{\mathrm{t}})$ in Eq. (\ref{121}) which is a generalized Rayleigh quotient problem, and the optimal transmit beamforming vector can be derived as given in Eq. (\ref{71}). The proof of the optimal receive combining vector follows the similar procedures and here we omit them due to space limitation.	
\end{IEEEproof}

Based on the receive combining and transmit beamforming vectors given above, the uplink and downlink channel capacity can be rewritten as 
\begin{equation}
\label{eq.8}
{\mathrm{C}}_{\mathrm{up}}=\frac{\tau_{\mathrm{data}}^{\mathrm{up}}}{\tau} \mathbb{E}\left\{ \max \; \log _{2}(1+\widetilde{\mathbf{h}}^{\mathrm{H}} \mathbf{U}^{-1} \widetilde{\mathbf{h}})\right\},
\end{equation}
\begin{equation}
\label{eq.9}
{\mathrm{C}}_{\mathrm{down}}=\frac{\tau_{\mathrm{data}}^{\mathrm{down}}}{\tau} \mathbb{E}\left\{ \max \; \log _{2}(1+\widetilde{\mathbf{h}}^{\mathrm{H}} \mathbf{D}^{-1} \widetilde{\mathbf{h}})\right\}.
\end{equation}

By using the Eq. (2.2) in \cite{silverstein1995empirical} (see Appendix for details), Eqs. (\ref{eq.8}) and (\ref{eq.9}) can be transformed to the following equivalent form,
\begin{equation}
\label{eq.10}
{\mathrm{C}}_{\mathrm{up}}=\frac{\tau_{\mathrm{data}}^{\mathrm{up}}}{\tau} \mathbb{E}\left\{ \max \; \log _{2}\left(1 + \frac{\widetilde{\mathbf{h}}^{\mathrm{H}} \widetilde{\mathbf{U}}^{-1} \widetilde{\mathbf{h}}}{1+\kappa_{\mathrm{UE}}\widetilde{\mathbf{h}}^{\mathrm{H}} \widetilde{\mathbf{U}}^{-1} \widetilde{\mathbf{h}}}\right)\right\},
\end{equation}
\begin{equation}
\label{eq.11}
{\mathrm{C}}_{\mathrm{down}}=\frac{\tau_{\mathrm{data}}^{\mathrm{down}}}{\tau} \mathbb{E}\left\{ \max \; \log _{2}\left(1 + \frac{\widetilde{\mathbf{h}}^{\mathrm{H}} \widetilde{\mathbf{D}}^{-1}\widetilde{\mathbf{h}}}{1+\kappa_{\mathrm{UE}}\widetilde{\mathbf{h}}^{\mathrm{H}} \widetilde{\mathbf{D}}^{-1} \widetilde{\mathbf{h}}}\right)\right\},
\end{equation}
where $ \widetilde{\mathbf{U}}=\left(1+\kappa_{\mathrm{UE}}\right) \kappa_{\mathrm{BS}} \operatorname{diag}(\widetilde{\mathbf{h}}\widetilde{\mathbf{h}}^{\mathrm{H}}) + \frac{\sigma_{\mathrm{UE}}^{2}}{p_{\mathrm{UE}}} \mathbf{I}$, and $ \widetilde{\mathbf{D}}=\left(1+\kappa_{\mathrm{UE}}\right) \kappa_{\mathrm{BS}} \operatorname{diag}(\widetilde{\mathbf{h}}\widetilde{\mathbf{h}}^{\mathrm{H}})+ \frac{\sigma_{\mathrm{UE}}^{2}}{p_{\mathrm{BS}}} \mathbf{I}$.

The maximizing operations in Eqs. (\ref{eq.10}) and (\ref{eq.11}) act on the IRS, \textit{i.e.}, optimize the phase shift parameters of the IRS matrix. Some optimization solutions for the cases with ideal hardware have been proposed in prior works \cite{8683145,8930608}. The hardware impairments will increase the complexity and difficulty of optimization. However, the optimization of IRS will not be discussed in this paper since what we want to obtain is the asymptotic channel capacity which will not be affected by the phase shifts of IRS. We consider two types of asymptotic channel capacity: the capacity as $ p_{\mathrm{BS}}\rightarrow \infty $ ($ p_{\mathrm{UE}}\rightarrow \infty $) and the capacity as $ M,N \rightarrow \infty $. In what follows, we only provide the result for the downlink due to space limitation. 

\begin{theorem}
	The asymptotic capacity limit $ \mathrm{C}^{p_{\mathrm{BS}}}_{\mathrm{down}}(\infty) = \lim _{p_{\mathrm{BS}} \rightarrow \infty}\mathrm{C}_{\mathrm{down}}$ is finite and bounded as 
	\begin{equation}
	\label{eq.12}
	\begin{aligned}
	\mathrm{C}^{p_{\mathrm{BS}}}_{\mathrm{down}}(\infty) &\leq \frac{\tau_{\mathrm{data}}^{\mathrm{down}}}{\tau} \log _{2}\left(1+\frac{M}{ \kappa_{\mathrm{BS}}+ \kappa_{\mathrm{UE}}(M+\kappa_{\mathrm{BS}})}\right), \\
	\mathrm{C}^{p_{\mathrm{BS}}}_{\mathrm{down}}(\infty) &\geq \frac{\tau_{\mathrm{data}}^{\mathrm{down}}}{\tau} \log _{2}\left(1+\frac{1}{ \kappa_{\mathrm{BS}}+ \kappa_{\mathrm{UE}}(1+\kappa_{\mathrm{BS}})}\right).
	\end{aligned}
	\end{equation}	
\end{theorem}

\begin{IEEEproof}
	The matrix $\widetilde{\mathbf{D}}$ is a diagonal matrix and the channel vector $\widetilde{\mathbf{h}}$ can be equivalently expressed as $ \mathbf{h}_{\mathrm{d}}+\mathbf{H}_{\mathrm{IRS}}\widetilde{\boldsymbol{\theta}} $ where $\widetilde{\boldsymbol{\theta}}$ is the column vector transformed from the IRS matrix $ \widetilde{\mathbf{\Phi}} $. Thus, $ \widetilde{\mathbf{h}}^{\mathrm{H}} \widetilde{\mathbf{D}}^{-1} \widetilde{\mathbf{h}} $ can be rewritten as 
	\begin{equation}
	\label{eq.13}
	\sum_{i=1}^{M} \frac{\left|{h}_{\mathrm{d}, i}+\mathbf{row}_{i} \widetilde{\boldsymbol{\theta}}\right|^{2}}{\left(1+\kappa_{\mathrm{UE}}\right)\kappa_{\mathrm{BS}}\left|{h}_{\mathrm{d}, i}+\mathbf{row}_{i} \widetilde{\boldsymbol{\theta}}\right|^{2}+\frac{\sigma_{\mathrm{BS}}^{2}}{p_{\mathrm{BS}}}},
	\end{equation}
	where $ {h}_{\mathrm{d}, i} $ is the $ i^{th} $ element of the direct channel vector $ \mathbf{h}_{\mathrm{d}} $, and  $ \mathbf{row}_{i} $ is the $ i^{th} $ row vector of the cascaded channel matrix $ \mathbf{H}_{\mathrm{IRS}} $. We have 
	\begin{equation}
	\label{eq.14}
	\lim _{p_{\mathrm{BS}} \rightarrow \infty} \mathbb{E}\left\{\widetilde{\mathbf{h}}^{\mathrm{H}} \widetilde{\mathbf{D}}^{-1} \widetilde{\mathbf{h}}\right\} = \frac{M}{\left(1+\kappa_{\mathrm{UE}}\right)\kappa_{\mathrm{BS}}}.
	\end{equation}
	By using Jensen's inequality in Eq. (\ref{eq.11}) and substituting Eq. (\ref{eq.14}) into it, we obtain the upper bound. The lower bound is asymptotically obtained by using $ \mathbb{E}\left\{\mathbf{x} \mathbf{x}^{\mathrm{H}}\right\} = \frac{p_{\mathrm{BS}}}{M} \mathbf{I}$.
\end{IEEEproof}

\begin{theorem}
	The asymptotic capacity limit $\mathrm{C}^{M,N}_{\mathrm{down}}(\infty) = \lim _{M,N \rightarrow \infty} \mathrm{C}_{\mathrm{down}} $ is finite and bounded as 
	\begin{equation}
	\label{eq.15}
	\begin{aligned}
	\mathrm{C}^{M,N}_{\mathrm{down}}(\infty) &\leq \frac{\tau_{\mathrm{data}}^{\mathrm{down}}}{\tau} \log _{2}\left(1+\frac{1}{\kappa_{\mathrm{UE}}}\right), \\
	\mathrm{C}^{M,N}_{\mathrm{down}}(\infty) &\geq \frac{\tau_{\mathrm{data}}^{\mathrm{down}}}{\tau} \log _{2}\left(1+\frac{1}{ \kappa_{\mathrm{BS}}+ \kappa_{\mathrm{UE}}(1+\kappa_{\mathrm{BS}})}\right).
	\end{aligned}
	\end{equation}
\end{theorem}
\begin{IEEEproof}
	Based on the form of $\widetilde{\mathbf{h}}^{\mathrm{H}} \widetilde{\mathbf{D}}^{-1}\widetilde{\mathbf{h}}$ in Eq. (\ref{eq.13}), we observe that $ \widetilde{\mathbf{h}}^{\mathrm{H}}\widetilde{\mathbf{D}}^{-1}\widetilde{\mathbf{h}}$ will approach to infinity as $ M,N \rightarrow \infty $. Thus, the upper bound of channel capacity converges to $ \frac{\tau_{\mathrm{data}}^{\mathrm{down}}}{\tau} \log _{2}\left(1+\frac{1}{\kappa_{\mathrm{UE}}}\right) $ as $ M,N \rightarrow \infty $. The lower bound is asymptotically achieved by using $\mathbb{E}\left\{\mathbf{x} \mathbf{x}^{\mathrm{H}}\right\} = \frac{p_{\mathrm{BS}}}{M} \mathbf{I}.$
\end{IEEEproof}

\begin{remark}
	When the number of BS antennas is finite and the number of reflecting elements approaches infinity, we see that the upper bound of asymptotic channel capacity is same as that of the case where $ p_{\mathrm{BS}} $ approaches infinity. This implies that the major function of IRS is to increase the power of  received signals, \textit{i.e.}, reach the channel capacity limit at low transmit power. That is also the reason why IRS can improve the energy and spectrum efficiency of wireless communication systems without the need of complex processing on signals. However, we cannot increase the asymptotic channel capacity assisted by the IRS at a certain $ M $.    
\end{remark}

\subsection{Energy Efficiency of Downlink}
In this subsection, we analyze the energy efficiency which is measured in Bit/Joule, and a common definition is the ratio of the spectral efficiency (in Bit/Channel Use) to the transmit power (in Joule/Channel Use). The energy consumed in the BS and the user (per coherence period) is 
\begin{equation}
\mathrm{E}=\tau_{\mathrm{data}}^{\mathrm{down}} p_{\mathrm{BS}} + \left(\tau_{\mathrm{pilot}}+ \tau_{\mathrm{data}}^{\mathrm{up}}\right) p_{\mathrm{UE}}.
\end{equation}
In addition, there exists a baseband circuit power consumption which can be modeled as $ M\rho + \zeta $ \cite{6056691,6251827}. The parameter $ \rho \geq 0 $ describes the circuit power which scales with the number of antennas $M$. The parameter $ \zeta > 0 $ describes the circuit power which is static. Then, the average power (in Joule/Channel Use) can be given as 
\begin{equation}
\label{eq.17}
\begin{aligned}
\frac{\mathrm{E}}{\tau} =  & \left( \frac{\tau_{\mathrm{data}}^{\mathrm{down}}}{\tau_{\mathrm{data}}^{\mathrm{up}}+\tau_{\mathrm{data}}^{\mathrm{down}}} \left(  \frac{\tau_{\mathrm{pilot}}p_{\mathrm{UE}}}{\tau} + M\rho + \zeta\right)  + \frac{\tau_{\mathrm{data}}^{\mathrm{down}} p_{\mathrm{BS}}}{\tau} \right)  \\ 
+& \left( \frac{\tau_{\mathrm{data}}^{\mathrm{up}}}{\tau_{\mathrm{data}}^{\mathrm{up}}+\tau_{\mathrm{data}}^{\mathrm{down}}} \left(  \frac{\tau_{\mathrm{pilot}}p_{\mathrm{UE}}}{\tau} + M\rho + \zeta \right) +  \frac{\tau_{\mathrm{data}}^{\mathrm{up}} p_{\mathrm{UE}}}{\tau} \right).
\end{aligned}
\end{equation}
The first term of Eq. (\ref{eq.17}) refers to the average power of the downlink transmission. Based on the power consumption modeled above, we give the definition of the overall energy efficiency as follows.

\begin{definition}
	The energy efficiency of downlink is 
	\begin{equation}
	\label{eq.18}
	\Xi_{\mathrm{down}} = \frac{\mathrm{C}_{\mathrm{down}}}{\frac{\tau_{\mathrm{data}}^{\mathrm{down}}}{\tau_{\mathrm{data}}^{\mathrm{up}}+\tau_{\mathrm{data}}^{\mathrm{down}}}\left(\frac{\tau_{\mathrm{pilot}}p_{\mathrm{UE}}}{\tau} + M\rho + \zeta \right) + \frac{\tau_{\mathrm{data}}^{\mathrm{down}} p_{\mathrm{BS}}}{\tau}},
	\end{equation}
	where $ \mathrm{C}_{\mathrm{down}} $ is the channel capacity of downlink.
\end{definition}

\begin{corollary}
	Suppose we want to maximize the energy efficiency of downlink with respect to the transmit power ($p_{\mathrm{BS}},p_{\mathrm{UE}} \geq 0$), the number of antennas ($M \geq 0 $), and the number of reflecting elements ($N \geq 0$). If the parameter $ \rho=0 $, the maximal energy efficiency $\Xi^{\mathrm{max}}_{\mathrm{down}}$ is bounded as 
	\begin{equation}
	\label{eq.19}
	\frac{ \log _{2}\left(1+\frac{1}{ \kappa_{\mathrm{BS}}+ \kappa_{\mathrm{UE}}(1+\kappa_{\mathrm{BS}})}\right)}{\frac{\tau\zeta}{\tau_{\mathrm{data}}^{\mathrm{up}}+\tau_{\mathrm{data}}^{\mathrm{down}}}}\leq  \Xi^{\mathrm{max}}_{\mathrm{down}} \leq \frac{\log _{2}\left(1+\frac{1}{\kappa_{\mathrm{UE}}}\right)}{\frac{\tau\zeta}{\tau_{\mathrm{data}}^{\mathrm{up}}+\tau_{\mathrm{data}}^{\mathrm{down}}}},
	\end{equation}
	If the parameter $ \rho>0 $, the upper bound is still valid, but the asymptotic energy efficiency is zero,\textit{ i.e.}, the maximal energy efficiency can be achieved at certain finite $ M $.
\end{corollary}

\begin{IEEEproof}
	Based on the definition given in Definition 1 and the asymptotic capacity of downlink in Theorem 3, we can prove Corollary 1. We maximize the energy efficiency with respect to the transmit powers, the number of antennas, and the number of reflecting elements: 1) by neglecting the transmit power terms in the denominator of Eq. (\ref{eq.18});  2) and applying the asymptotic capacity limit bounds in Theorem 3 to the numerator of Eq. (\ref{eq.18}). Then, we obtain the upper bound and the lower bound of energy efficiency of downlink in Eq. (\ref{eq.19}).
\end{IEEEproof}

\begin{remark}
	If the number of BS antennas is fixed and the circuit power have no correlation with the number of antennas at BS, \textit{i.e.}, $ \rho=0 $, increasing the number of reflecting elements on IRS can only make the energy efficiency close to the upper bound given as follows,
	\begin{equation}
	\label{eq.20}
	\Xi^{\mathrm{max}}_{\mathrm{down}} \leq \frac{\log _{2}\left(1+\frac{M}{ \kappa_{\mathrm{BS}}+ \kappa_{\mathrm{UE}}(M+\kappa_{\mathrm{BS}})}\right)}{\frac{\tau\zeta}{\tau_{\mathrm{data}}^{\mathrm{up}}+\tau_{\mathrm{data}}^{\mathrm{down}}}}.
	\end{equation}
	The upper bound of maximal energy efficiency only can be increased from Eq. (\ref{eq.20}) to Eq. (\ref{eq.19}) by increasing the number of BS antennas.
	
	If the circuit power scales with the number of BS antennas, \textit{i.e.}, $ \rho>0 $, the maximal energy efficiency only can be achieved at some finite $ M $, which depends on the parameters $ \rho $ and $ \zeta $. Meanwhile, the effect of increasing the reflecting elements of IRS in this case is still to make the energy efficiency close to the upper bound under the current $ M $.
\end{remark}

\section{Numerical Results}
\begin{figure}[htbp]
	\centerline{\includegraphics[width=8.5 cm]{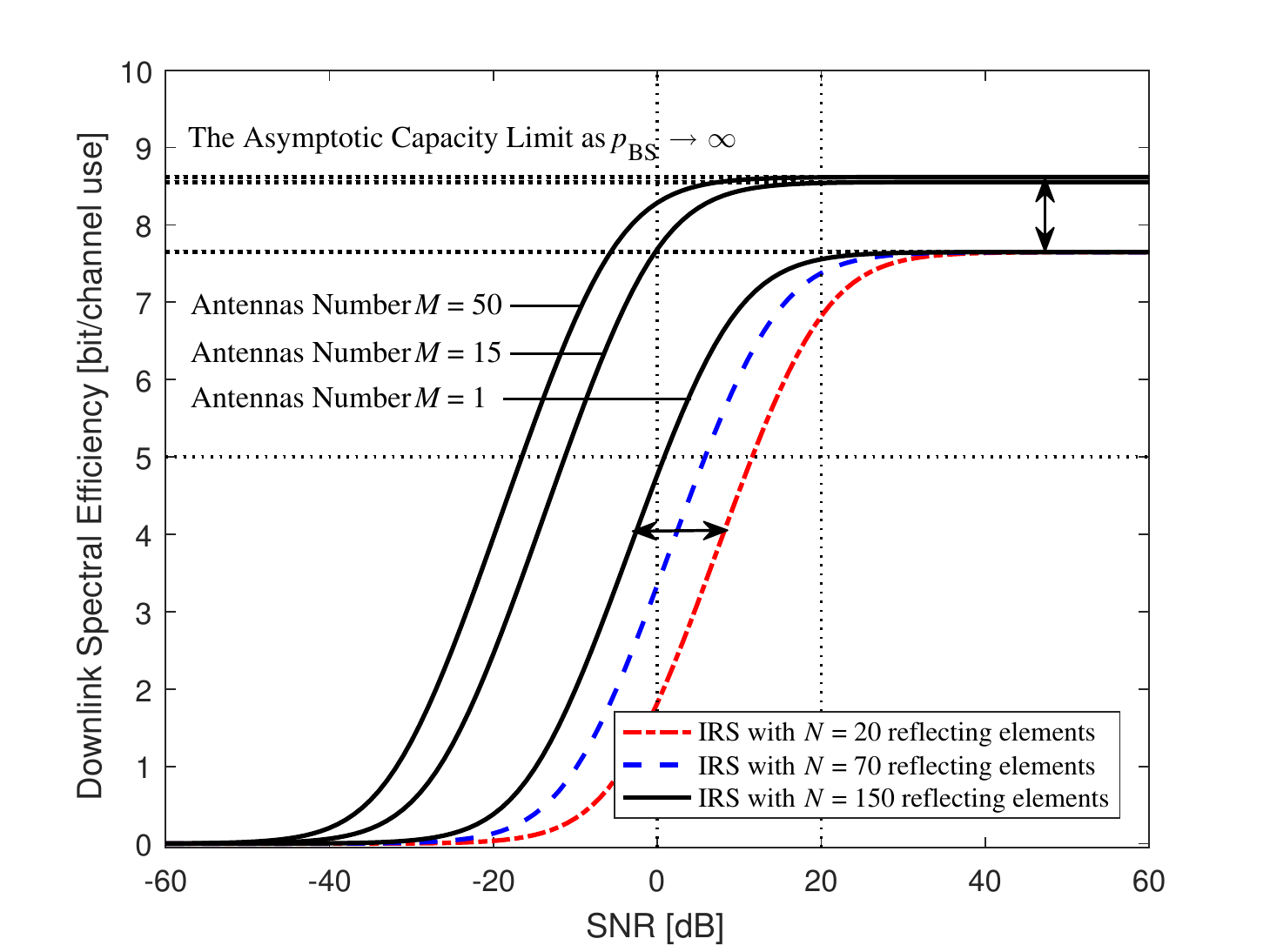}}
	\caption{Spectral efficiency on the downlink versus SNR for $ \kappa_{\mathrm{BS}} = \kappa_{\mathrm{UE}} = 0.05^2 $ with different numbers of BS antennas and reflecting elements.}
	\label{fig3}
	\vspace{-1.2 em}
\end{figure}

In this section, we numerically illustrate the results proposed in Section III. To evaluate the asymptotic channel capacity $\mathrm{C}^{p_{\mathrm{BS}}}_{\mathrm{down}}(\infty) = \lim _{p_{\mathrm{BS}} \rightarrow \infty}\mathrm{C}_{\mathrm{down}}$, especially the upper bound of it, we consider the number of BS antennas $ M $ in the set of $ \left\lbrace 1, 15, 50 \right\rbrace  $ and the number of reflecting elements on IRS $N$ in the set of $ \left\lbrace 20, 70, 150 \right\rbrace $. The hardware impairments coefficients are set as $ \kappa_{\mathrm{BS}} = \kappa_{\mathrm{UE}} = 0.05^2 $. The optimization of IRS will not be considered in this simulation since the asymptotic channel capacity will not be affected by the phase shifts of IRS. Thus, we set the matrix of IRS as an identity matrix with the size of $ N \times N $. Fig.~\ref{fig3} shows the spectral efficiency on downlink versus the SNR with different numbers of BS antennas and reflecting elements. The SNR is defined as $ p_{\mathrm{BS}}/\sigma^2_{\mathrm{UE}} $. It is observed that as $ p_{\mathrm{BS}} \rightarrow \infty $, the asymptotic channel capacity will converge to a finite value which is related to the hardware impairments and the number of BS antennas $ M $. When we fix $ N = 150 $ and raise $ M $ from 1 to 15, there is a significant increase in the asymptotic channel capacity limit. However, when we raise $ M $ substantially from 15 to 50, the growth of the asymptotic channel capacity limit becomes less noticeable. This implies that there is no need to increase the number of BS antennas when it reaches a certain number. When we fix $M = 1$ and assign the value of $N$ in the set of $ \left\lbrace 20, 70, 150 \right\rbrace $, we obtain that higher spectral efficiency at a low SNR can be achieved with more reflecting elements.

\begin{figure}[htbp]
	\centerline{\includegraphics[width=8.5 cm]{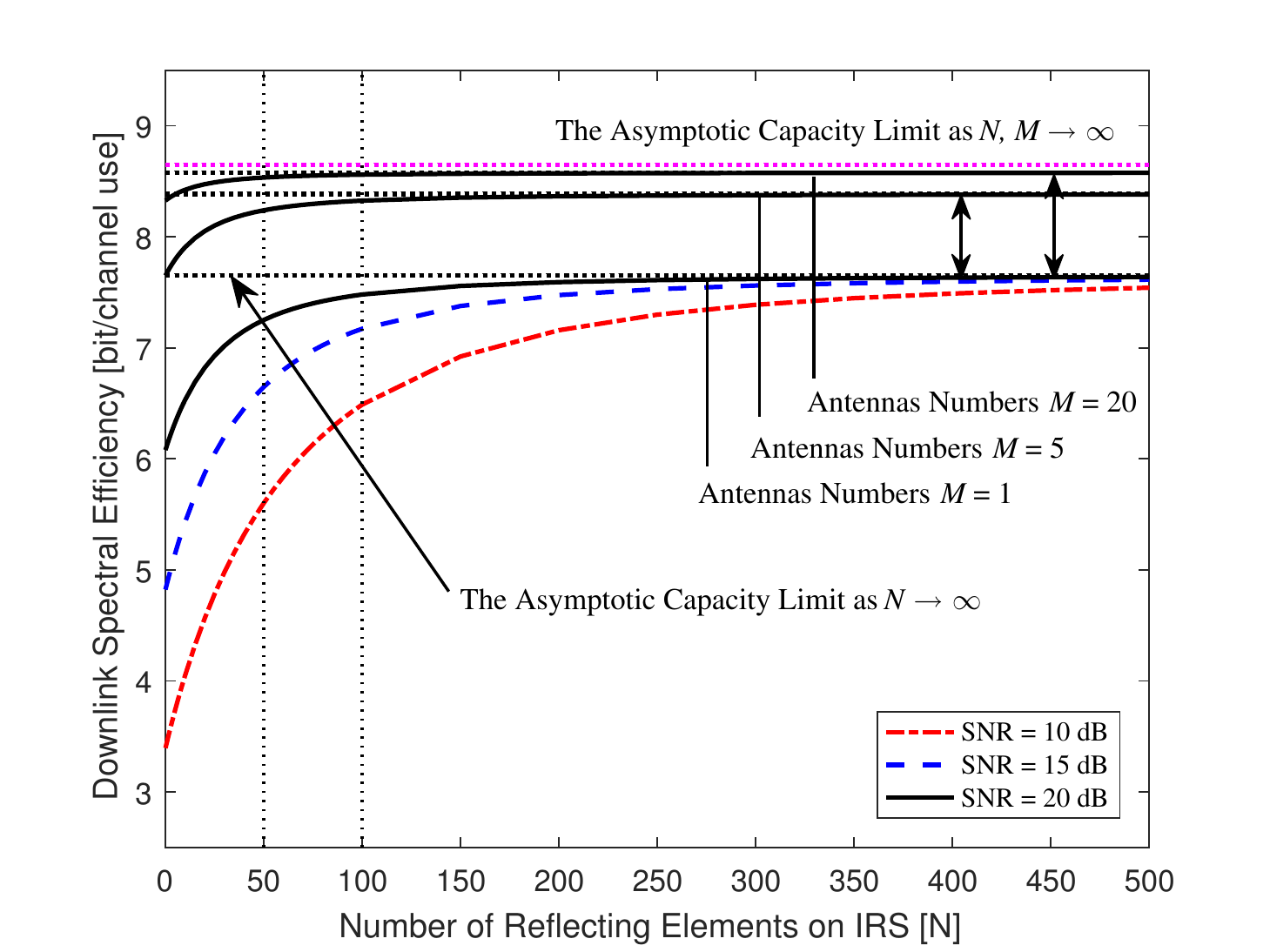}}
	\caption{Spectral efficiency on the downlink versus the number of reflecting elements on IRS with different transmit power and BS antennas.}
	\label{fig4}
	\vspace{-0.2 em}
\end{figure}

To evaluate the asymptotic channel capacity $\mathrm{C}^{M,N}_{\mathrm{down}}(\infty) = \lim _{M,N \rightarrow \infty} \mathrm{C}_{\mathrm{down}}$, especially the upper bound of it, we consider the SNR, which is defined as $p_{\mathrm{BS}}/\sigma^2_{\mathrm{UE}}$, in the set of $\left\lbrace 10\text{ dB}, 15\text{ dB}, 20\text{ dB} \right\rbrace $ and the number of BS antennas $M$ in the set of $ \left\lbrace 1, 5, 20 \right\rbrace $. We set the hardware impairments coefficients as $ \kappa_{\mathrm{BS}} = \kappa_{\mathrm{UE}} = 0.05^2 $, and set the matrix of IRS as an identity matrix with the size of $ N \times N $. Fig.~\ref{fig4} shows the spectral efficiency on downlink versus the number of reflecting elements on IRS with different transmit power and different numbers of BS antennas. It is shown that as $ M, N \rightarrow \infty $, the asymptotic channel capacity will converge to a finite value $ \log _{2}\left(1+\frac{1}{\kappa_{\mathrm{UE}}}\right) $ which is related to the hardware impairments. When we fix $ M = 1 $, the spectral efficiency  will converge to a finite number as $ N \rightarrow \infty $, which is same as the case of $ p_{\mathrm{BS}} \rightarrow \infty $ in Fig.~\ref{fig3}. This confirms that the major effect of IRS is to increase the power of  received signals, \textit{i.e.}, reach the channel capacity limit at low transmit power. When we raise $ p_{\mathrm{BS}}/\sigma^2_{\mathrm{UE}} $ from 10 dB to 20dB, the same spectral efficiency can be achieved with less number of reflecting elements. When we raise $ M $ from 1 to 5, there is a significant increase in the asymptotic channel capacity limit. However, when we raise $ M $ substantially from 5 to 20, the growth of the asymptotic channel capacity limit becomes less noticeable, and is very close to $ \log _{2}\left(1+\frac{1}{\kappa_{\mathrm{UE}}}\right) $.

\vspace{-1.2 em}

\vspace{-1.3 em}

\begin{figure}[htbp]
	\centerline{\includegraphics[width=8.5 cm]{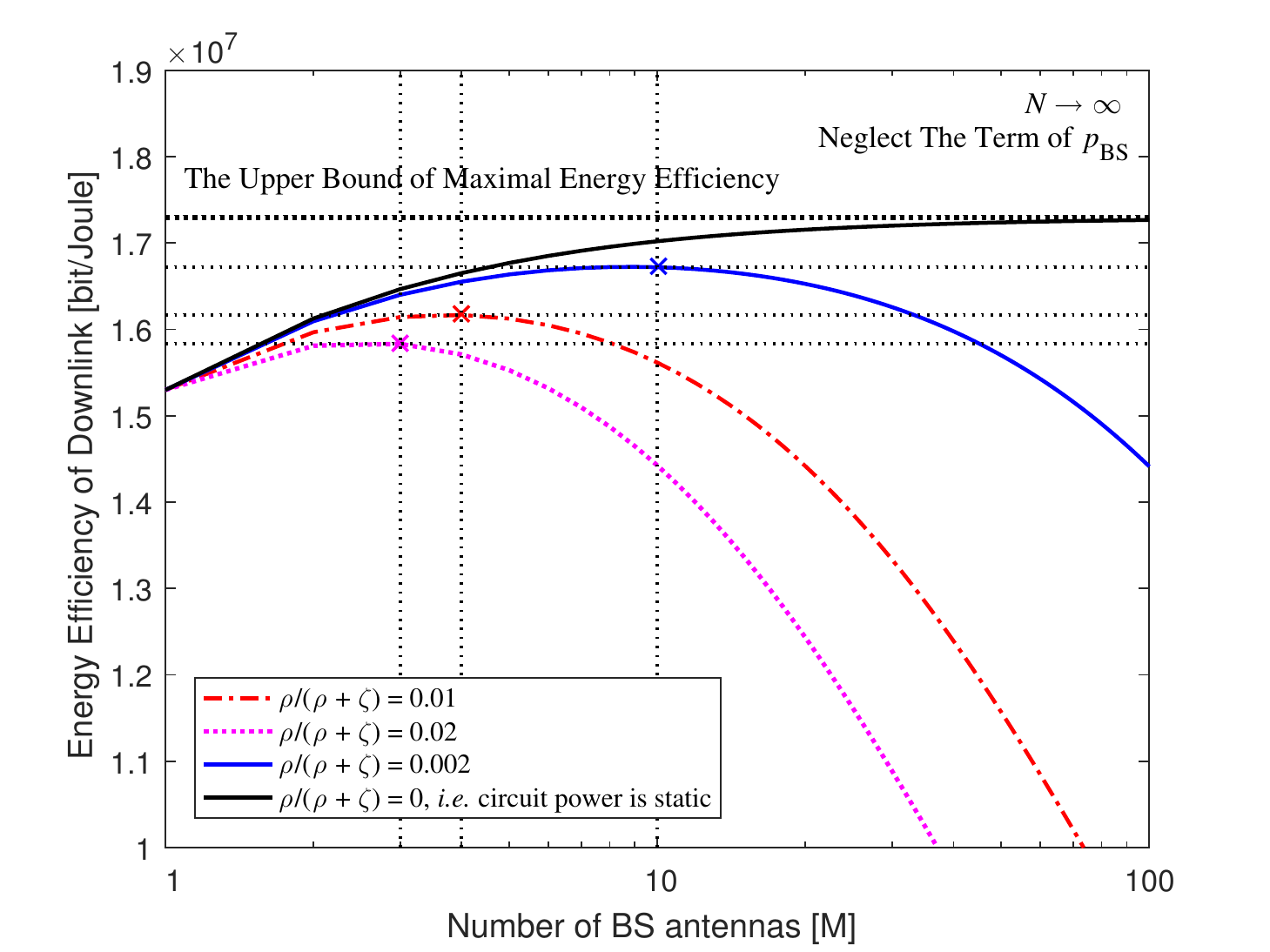}}
	\caption{Maximal energy efficiency of downlink versus the number of antennas with different splitting between $ \rho $ and $ \zeta $. In this case, we set the number of reflecting elements $ N \rightarrow \infty $, and neglect the transmit power terms.  }
	\label{fig5}
	\vspace{-1 em}
\end{figure}

To evaluate the maximal energy efficiency of IRS-assisted system, we consider the case where the number of reflecting elements $N$ approaches to infinity. In this case, the upper bound of asymptotic channel capacity is same as that of the case where the transmit power $p_{\mathrm{BS}}$ approaches to infinity. Thus, we neglect the transmit power terms in the energy efficiency. It is reasonable because the energy efficiency can be improved by reducing the transmit power without the reduction of spectral efficiency. To illustrate the difference between the static circuit power and the circuit power which scales with the number of BS antennas, we consider four different splittings between $ \rho $ and $ \zeta $: $ \frac{\rho}{\rho+\zeta} \in \{0, 0.002, 0.01, 0.02\} $. Based on the power consumption numbers reported in \cite{auer2010d2}, we consider the power consumption that $ \rho+\zeta = 0.5\mu \; \text{Joule/Channel Use} $. Fig.~\ref{fig5} shows the maximal energy efficiency versus the number of BS antennas with different splitting between $ \rho $ and $ \zeta$. It is seen that when $ \rho = 0 $, \textit{i.e.}, the circuit power is static, the maximal energy efficiency increases with the number of BS antennas and converge to a finite value, which conforms to the upper bound proven in Corollary 1. While $ \rho > 0 $, \textit{i.e.}, some part of power consumption scales with $ M $, the maximal energy efficiency can be achieved at some finite $ M $. When the part of power consumption which scales with $ M $ is non-negligible, \textit{e.g.}, $\frac{\rho}{\rho+\zeta} = 0.01, 0.02$ in Fig.~\ref{fig5}, the number of BS antennas is not necessary to be very large. This conclusion is similar to the results in Fig.~\ref{fig3} and Fig.~\ref{fig4}. This implies that an IRS-assisted wireless communication system can achieve both high spectral efficiency and high energy efficiency with small number of BS antennas. As to the number of reflecting elements on IRS, it should be as large as possible without degrading the performance, \textit{e.g.}, channel estimation accuracy, bit error rate. This conclusion is encouraging as there is no need to cost a lot on expensive high-quality antennas when the system is assisted by an IRS, which corresponds to the requirements of new communication paradigms.

\vspace{-1.2 em}

\section{Conclusion}
This paper analyzes the optimal receive combining and transmit beamforming vectors, the asymptotic channel capacity, and the energy efficiency of the IRS-assisted wireless communication system with hardware impairments. We show that although the hardware impairments increase the complexity and difficulty of reflect beamforming design, the asymptotic channel capacity is not affected by the phase shifts of IRS. It is also shown that an IRS-assisted wireless communication system can achieve both high spectral efficiency and high energy efficiency with moderate number of antennas. This result is encouraging for there is no need to cost a lot on expensive high-quality antennas, which verifies the requirements of new communication paradigms.

\vspace{-0.1 em}
\appendix{The Equation (2.2) in Reference \cite{silverstein1995empirical}: }
For $ N \times N $ matrix $ \mathbf{B} $, $ \tau \in \mathbb{C} $ and $ \mathbf{q} \in \mathbb{C}^{N} $ for which $ \mathbf{B} $ and $ \mathbf{B} + \tau\mathbf{q}\mathbf{q}^{\mathrm{H}} $ are invertible, it holds that
\begin{equation}
\mathbf{q}^{\mathrm{H}} (\mathbf{B} + \tau\mathbf{q}\mathbf{q}^{\mathrm{H}})^{-1} = \frac{1} {1+\tau\mathbf{q}^{\mathrm{H}}\mathbf{B}^{-1}\mathbf{q}} \mathbf{q}^{\mathrm{H}} \mathbf{B}^{-1}. 
\end{equation}

\balance

\bibliographystyle{IEEEtran} 
\bibliography{reference}

\begin{thebibliography}{10}
\providecommand{\url}[1]{#1}
\csname url@samestyle\endcsname
\providecommand{\newblock}{\relax}
\providecommand{\bibinfo}[2]{#2}
\providecommand{\BIBentrySTDinterwordspacing}{\spaceskip=0pt\relax}
\providecommand{\BIBentryALTinterwordstretchfactor}{4}
\providecommand{\BIBentryALTinterwordspacing}{\spaceskip=\fontdimen2\font plus
\BIBentryALTinterwordstretchfactor\fontdimen3\font minus
  \fontdimen4\font\relax}
\providecommand{\BIBforeignlanguage}[2]{{%
\expandafter\ifx\csname l@#1\endcsname\relax
\typeout{** WARNING: IEEEtran.bst: No hyphenation pattern has been}%
\typeout{** loaded for the language `#1'. Using the pattern for}%
\typeout{** the default language instead.}%
\else
\language=\csname l@#1\endcsname
\fi
#2}}
\providecommand{\BIBdecl}{\relax}
\BIBdecl

\bibitem{index2019global}
\BIBentryALTinterwordspacing
``Cisco annual internet report (2018-2023),'' Mar. 2020. [Online]. Available:
  \url{https://www.cisco.com/c/en/us/solutions/collateral/executive-perspectives/annual-internet-report/white-paper-c11-741490.pdf}
\BIBentrySTDinterwordspacing

\bibitem{8796365}
E.~{Basar}, M.~{Di Renzo}, J.~{De Rosny}, M.~{Debbah}, M.~{Alouini}, and
  R.~{Zhang}, ``Wireless communications through reconfigurable intelligent
  surfaces,'' \emph{IEEE Access}, vol.~7, pp. 116\,753--116\,773, 2019.

\bibitem{8930608}
Q.~{Wu} and R.~{Zhang}, ``Beamforming optimization for wireless network aided
  by intelligent reflecting surface with discrete phase shifts,'' \emph{IEEE
  Transactions on Communications}, vol.~68, no.~3, pp. 1838--1851, Mar. 2020.

\bibitem{8936989}
{\"O}.~{{\"O}zdogan}, E.~{Björnson}, and E.~G. {Larsson}, ``Intelligent
  reflecting surfaces: Physics, propagation, and pathloss modeling,''
  \emph{IEEE Wireless Communications Letters (Eearly Access)}, pp. 1--1, 2019.

\bibitem{8647620}
Q.~{Wu} and R.~{Zhang}, ``Intelligent reflecting surface enhanced wireless
  network: Joint active and passive beamforming design,'' in \emph{2018 IEEE
  Global Communications Conference (GLOBECOM)}, Abu Dhabi, United Arab
  Emirates, United Arab Emirates, Dec. 2018, pp. 1--6.

\bibitem{6362131}
E.~{Bj{\"o}rnson}, P.~{Zetterberg}, M.~{Bengtsson}, and B.~{Ottersten},
  ``Capacity limits and multiplexing gains of {MIMO} channels with transceiver
  impairments,'' \emph{IEEE Communications Letters}, vol.~17, no.~1, pp.
  91--94, Jan. 2013.

\bibitem{6457363}
H.~Q. {Ngo}, E.~G. {Larsson}, and T.~L. {Marzetta}, ``Energy and spectral
  efficiency of very large multiuser {MIMO} systems,'' \emph{IEEE Transactions
  on Communications}, vol.~61, no.~4, pp. 1436--1449, Apr. 2013.

\bibitem{nadeem2019intelligent}
Q.-U.-A. Nadeem, A.~Kammoun, A.~Chaaban, M.~Debbah, and M.-S. Alouini,
  ``Intelligent reflecting surface assisted multi-user {MISO} communication,''
  \emph{arXiv preprint arXiv:1906.02360}, 2019.

\bibitem{5456453}
C.~{Studer}, M.~{Wenk}, and A.~{Burg}, ``{MIMO} transmission with residual
  transmit-{RF} impairments,'' in \emph{2010 International ITG Workshop on
  Smart Antennas (WSA)}, Bremen, Germany, Feb. 2010, pp. 189--196.

\bibitem{zetterberg2011experimental}
P.~Zetterberg, ``Experimental investigation of {TDD} reciprocity-based
  zero-forcing transmit precoding,'' \emph{EURASIP Journal on Advances in
  Signal Processing}, vol. 2011, pp. 1--10, 2011.

\bibitem{8869792}
M.~{Badiu} and J.~P. {Coon}, ``Communication through a large reflecting surface
  with phase errors,'' \emph{IEEE Wireless Communications Letters}, vol.~9,
  no.~2, pp. 184--188, Feb. 2020.

\bibitem{8683145}
Q.~{Wu} and R.~{Zhang}, ``Beamforming optimization for intelligent reflecting
  surface with discrete phase shifts,'' in \emph{ICASSP 2019 - 2019 IEEE
  International Conference on Acoustics, Speech and Signal Processing
  (ICASSP)}, Brighton, United Kingdom, United Kingdom, May 2019, pp.
  7830--7833.

\bibitem{6891254}
E.~{Bj{\"o}rnson}, J.~{Hoydis}, M.~{Kountouris}, and M.~{Debbah}, ``Massive
  {MIMO} systems with non-ideal hardware: Energy efficiency, estimation, and
  capacity limits,'' \emph{IEEE Transactions on Information Theory}, vol.~60,
  no.~11, pp. 7112--7139, Nov. 2014.

\bibitem{silverstein1995empirical}
J.~W. Silverstein and Z.~Bai, ``On the empirical distribution of eigenvalues of
  a class of large dimensional random matrices,'' \emph{Journal of Multivariate
  analysis}, vol.~54, no.~2, pp. 175--192, 1995.

\bibitem{6056691}
G.~{Auer}, V.~{Giannini}, C.~{Desset}, I.~{Godor}, P.~{Skillermark},
  M.~{Olsson}, M.~A. {Imran}, D.~{Sabella}, M.~J. {Gonzalez}, O.~{Blume}, and
  A.~{Fehske}, ``How much energy is needed to run a wireless network?''
  \emph{IEEE Wireless Communications}, vol.~18, no.~5, pp. 40--49, Oct. 2011.

\bibitem{6251827}
D.~W.~K. {Ng}, E.~S. {Lo}, and R.~{Schober}, ``Energy-efficient resource
  allocation in {OFDMA} systems with large numbers of base station antennas,''
  \emph{IEEE Transactions on Wireless Communications}, vol.~11, no.~9, pp.
  3292--3304, Sept. 2012.

\bibitem{auer2010d2}
G.~Auer, O.~Blume, V.~Giannini, I.~Godor, M.~Imran, Y.~Jading, E.~Katranaras,
  M.~Olsson, D.~Sabella, P.~Skillermark \emph{et~al.}, ``D2. 3: Energy
  efficiency analysis of the reference systems, areas of improvements and
  target breakdown,'' \emph{Earth}, vol.~20, no.~10, 2010.

\end{thebibliography}
\end{document}